\documentclass[a4paper,11pt]{article}
\pdfoutput=1 

\usepackage{jheppub} 

\usepackage{siunitx}
\usepackage{multirow}
\newcommand{\ba}{\begin{eqnarray}}
\newcommand{\ea}{\end{eqnarray}}

\usepackage[T1]{fontenc} 

\title{\boldmath Lab at Home of fluid draining with Tracker}

\author[a]{Punsiri Dam-O,}
\author[a]{Thammarong Eadkong,}
\author[a,b]{Phongpichit Channuie}

\affiliation[a]{School of Science, Walailak University, Thasala, Nakhon Si Thammarat, 80160, Thailand}
\affiliation[b]{College of Graduate Studies, Walailak University, Thasala, Nakhon Si Thammarat, \\80160, Thailand}

\emailAdd{punsiri.pla@mail.wu.ac.th,thammarong.ea@mail.wu.ac.th,channuie@gmail.com,}

\abstract{The pandemic caused by the SARS-CoV-2 virus has threaten the face-to-face teaching activity in both schools and universities. The transformation from face-to-face classes to online activities yielded inefficient outcomes from the activities. Particularly, a key issue was organizing laboratory activities without accessing the labs. In this paper, we propose how to overcome this problem by enabling the students to perform physics experiments of fluid draining at home. In other words, the use of equipment commonly available at home or that can be purchased at a low price is practically plausible. In the present work, we do an experimental investigation of liquid draining through a
hole of a container. From an analyzing step, we introduce a freely-accessible video analysis software, Tracker, to obtain accurate results. Interestingly, we observe the effects of viscosity causing a delay of draining time of liquids. Our study shows that a ratio of open-space radius of the container to the hole radius, $\sqrt{\lambda}$, can be used to determine a draining time.}

\begin{document} 
\maketitle
\flushbottom

\section{Introduction}

The COVID-19 outbreak has disrupted face-to-face Physics lab course at university and has transformed them to online courses. It is challenging for physics instructors worldwide, in designing labs for undergraduate students to meet their proficiency. In late 2021, during the forth wave of COVID-19 of Thailand, we organized Physics Laboratory II course for engineering students. All 243 students had to study from home. Since previous course in Physics Laboratory I, all labs we served for our students were mainly rely on virtual labs, this second course we intended to arrange more hands-on labs. However The hands-on labs are based on the utilization of household items, similar to what have done by other universities \cite{labathome}.

One of our hands-on labs is exploration of the speed of efflux. In this lab, the students were assigned to calculate ratio of cross-sectional areas of a bottle to a leakage, and an initial height of water in the bottle. The students conduct the experiment with just ordinary equipment, e.g. a used transparent plastic bottle, a stopwatch and camera installed in their smartphone, a ruler, and tap water. Due to individual performing experiment, it prevents them from having precise time reckoning. As a result, a number of the students obtained significant errors in data analysis. To avoid significant errors in data analysis for the next academic year, we thus have introduced a VDO frame-based analysis software, called Tracker. In this proposed lad, we not only consider water but also choose oil as an another example of fluid into the lab. With an assist of Tracker and the use of different types of liquid, we hope that our students will be able to obtain more accurate data and to observe effects of density and viscosity to Bernoulli's and Poiseuille's laws. Before establishment of this new lab, we carried out a pilot experimental study to validate our method and accuracy of the experimental results. 

In this paper, we propose a new development of the speed of efflux lab. With a common plastic bottle and a free Tracker software, this lab enables students to conduct the experiment at home. Moreover, it brings more accurate measuring data to obtain a precise prediction of parameters such as the size of a small leakage hole and terminal time of fluid flow. In Sec.\ref{s2}, we revisit the mathematical formulation of the underlying description via the Bernoulli’s equation to describe the behavior of the flow. A vital component of our formal setup is given in Sec.\ref{s3}. Here it is rather straightforward, and helps students to establish the proper thought process. In Sec.\ref{s4}, we show the results and discuss our findings. Finally, we conclude in the last section. 

\section{Mathematical Formalism of Underlying Theory}\label{s2}
\begin{figure}[ht]
\begin{center}
\includegraphics[width=0.15\linewidth]{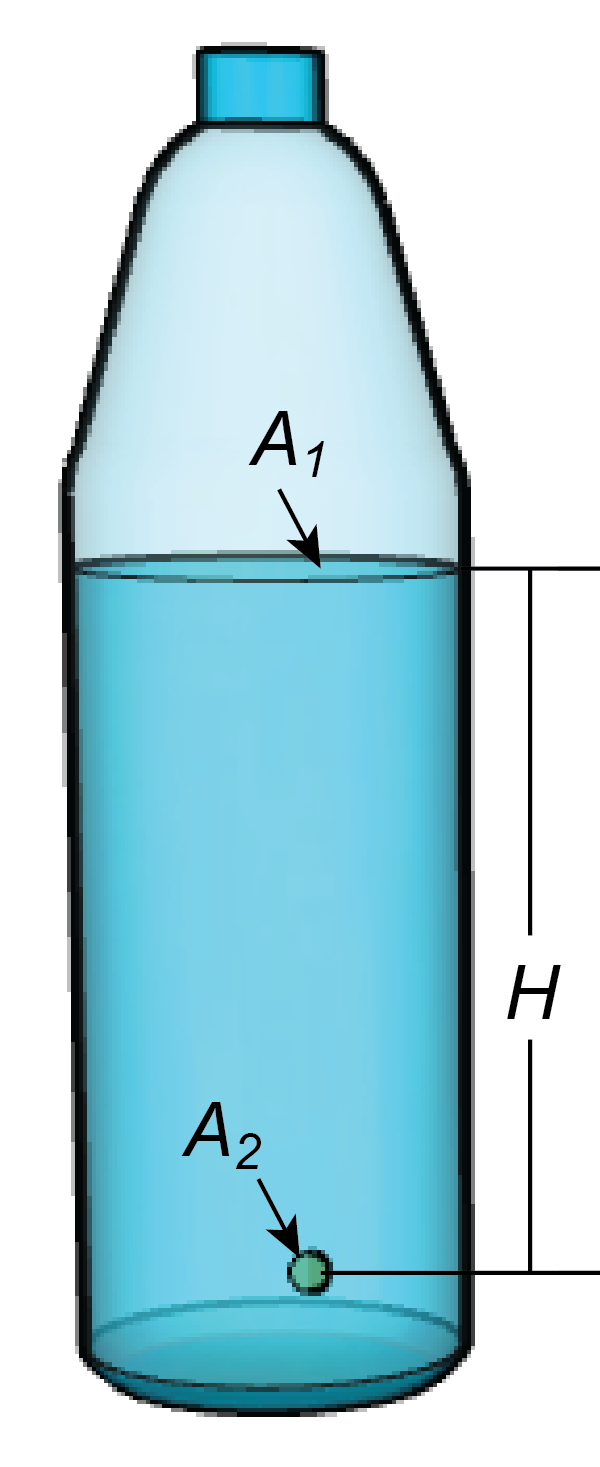} 
\caption{Schematic diagram of a bottle with the position of an orifice, the height of the fluid level $H$ between the free surface and an orifice, and cross-section areas of the bottle and the orifice.} 
\label{botle Fig} 
\end{center}
\end{figure}

Historically, Swiss physicist Daniel Bernoulli had first described the relationship between fluid speed, pressure, and elevation which is nowadays well known as the Bernoulli's equation \cite{Serway}. The equation plays a key role when one wants to describe the ideal fluid, i.e., the steady and in-compressible flow of non-viscous fluids.  We start considering the energy conservation of energy of the laminar flow. Firstly, we assume that the origin of coordinates is at the center of the orifice. As illustrated in Fig.\ref{botle Fig}, we label a position on the free surface of the liquid with \lq\lq\,1\,\rq\rq\,, and use \lq\lq\,2\,\rq\rq\, to represent a position of a hole in which the origin of coordinates is located.  Subsequently, we apply Bernoulli's equation to describe the behavior of the flow and write the total mechanical energy per volume as Eq.(\ref{str1}).  
\begin{eqnarray}
{P_{1}+\frac{1}{2}\rho v_{1}^{2}+\rho g y_{1}=P_{2}+\frac{1}{2}\rho v_{2}^{2}+\rho g y_{2}},\label{str1}
\end{eqnarray}
diswhere $P_{1}$ and $P_{2}$ represent the pressure at the free surface (1) and at an orifice (2), respectively, both equal to $P_{a}$, $\rho$ is the density of fluid which is constant. The free surface has the speed of $v_{1}$ and the speed of efflux is denoted by $v_{2}$, $g$ represents the gravitational acceleration. The height of the fluid vertically varies at the point 1 of an initial height of $H=H_{0}$ and at any time with $y_{1}$ as shown in Fig.{\ref{botle Fig}} and $y_{2}$ is zero which is the point of the origin of coordinates. Therefore, Eq.(\ref{str1}) reduces to 
\begin{eqnarray}
{\frac{1}{2}\rho v_{1}^{2}+\rho g H=\frac{1}{2}\rho v_{2}^{2}}.\label{str2}
\end{eqnarray}
For an ideal fluid flow that the volume flux is constant, we can write
\begin{eqnarray}
{A_{1}v_{1}=A_{2}v_{2}},\label{str3}
\end{eqnarray}
where $A_{1}$ and $A_{2}$ are cross-sectional areas of the bottle and the pierced hole, respectively. Now, we consider $v_{1}$ and recast it in terms of $v_{1}=dH/dt$. Substituting the result into Eq.(\ref{str2}), we obtain $v_{2}$:
\begin{eqnarray}
v_{2}=\sqrt{\bigg(\frac{dH}{dt}\bigg)^{2}+2gH}\,.\label{str31}
\end{eqnarray}
Therefore, from Eq.(\ref{str3}), we then have 
\begin{eqnarray}
{\frac{dH}{dt}=-\lambda \sqrt{\frac{2gH}{1-\lambda^2}}, \text{ or} \quad\quad \frac{dH}{\sqrt{H}}=-\lambda \sqrt{\frac{2g}{1-\lambda^2}}dt},\label{str4}
\end{eqnarray}
where $\lambda$ is denoted as a ratio of $A_{2}$ to $A_{1}$. A negative sign in front of Eq.(\ref{str4}) implies a decreasing the height ($H$) of the fluid in the container. Integrating out Eq.({\ref{str4}}) by using separation of variables, we have 
\begin{eqnarray}
H(t)=\frac{1}{2}\frac{g\lambda^2}{(1-\lambda^{2})} t^2 -\sqrt{\frac{2g\lambda^2}{(1-\lambda^{2})} H_{0}}\,t+H_{0},\label{str5}
\end{eqnarray}
where $H_{0}$ is the initial height of the free surface. If the cross-section area of the bottle is much greater than that of the area of the orifice ($A_{1}\gg A_{2}$), we find that $1-\lambda^2$ is equal to $1$ and Eq.(\ref{str5}) then becomes
\begin{eqnarray}
H(t)=\frac{1}{2}g\lambda^2 t^2 -\sqrt{2g\lambda^2 H_{0}}\,t+H_{0}.\label{str6}
\end{eqnarray}
In the present study, the above approximation will be verified by our experiments of fluid draining.

\section{Material \& Method}\label{s3}
\begin{figure}[ht]
\begin{center}
\includegraphics[width=0.7\linewidth]{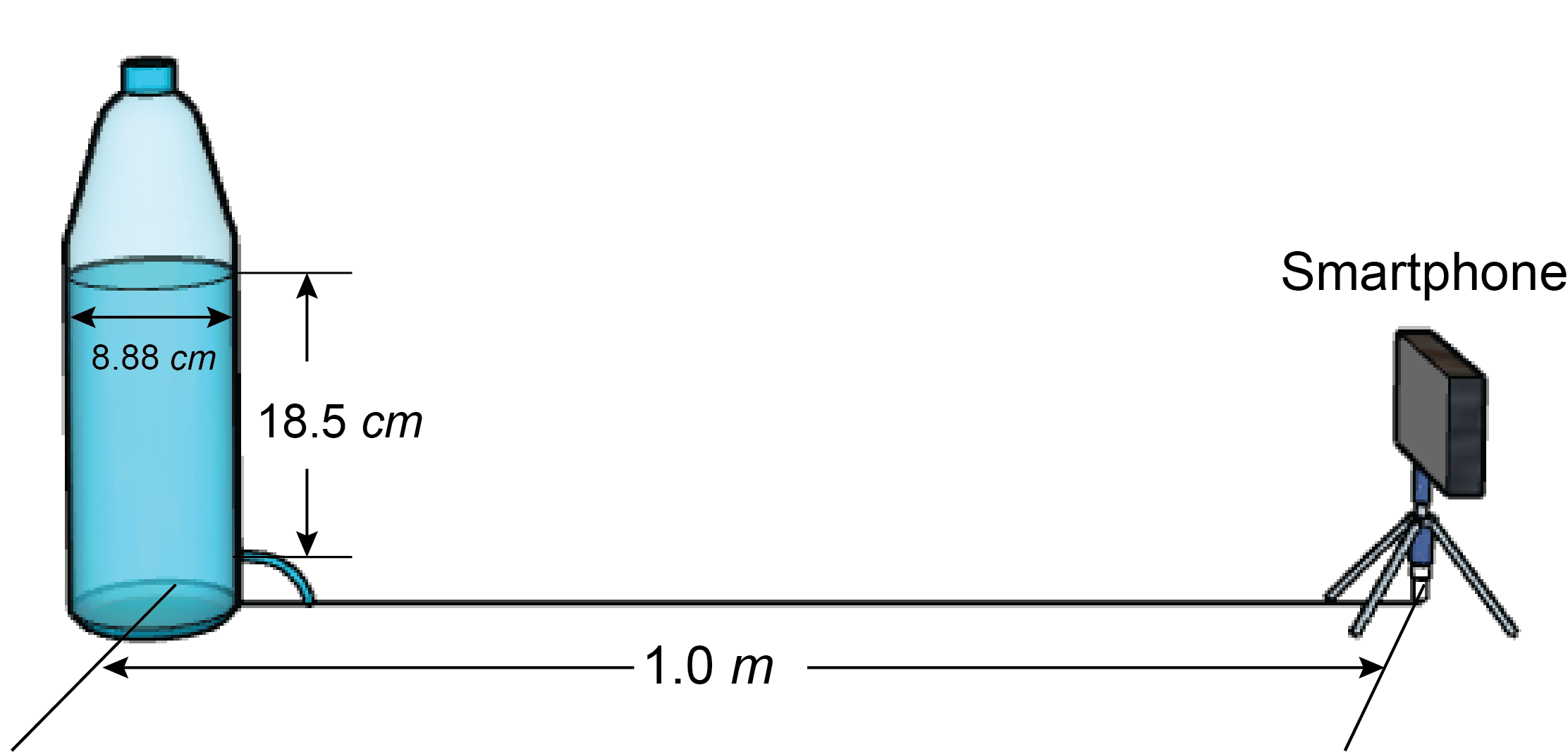} 
\caption{We use a smartphone to record the draining. An initial height of liquid is labelled with $H_{0}$.} 
\label{Setup}
\end{center}
\end{figure}

\begin{figure}[ht]
\begin{center}\includegraphics[width=0.8\linewidth]{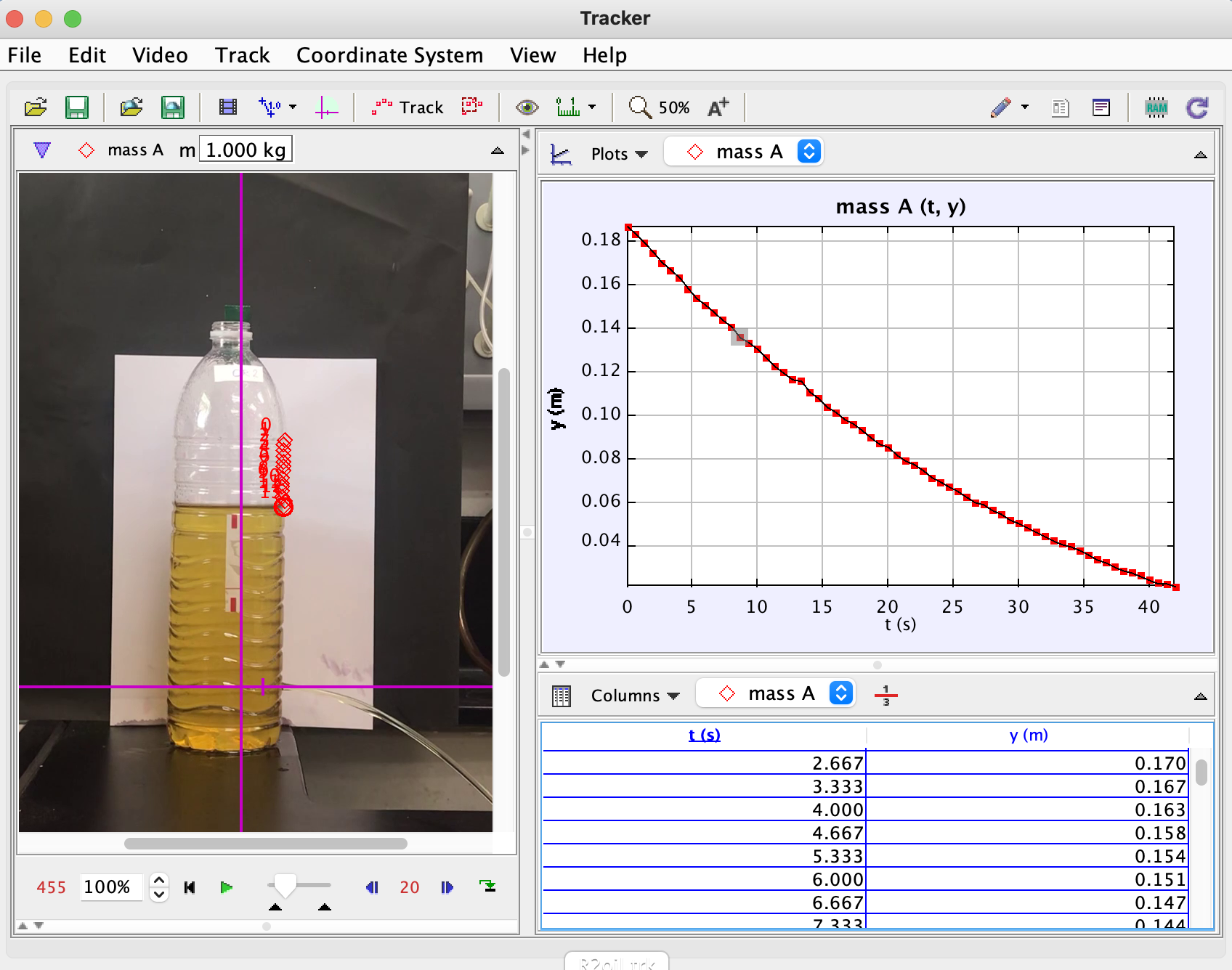}
\caption{The analysis panel consists of 3 portions: (left panel) a cross-section reference point of origin $(x=0,\,y=0)$, (upper-right panel) a plot of vertical positions of the fluid levels at different time points, and (bottom-right panel) a table of the vertical displacements at different time points.} 
\label{Tracker analysis}
\end{center}
\end{figure}

To study the speed of efflux using Bernoulli's equation, transparent plastic bottles--cylindrical shape were pierced by using a hot metal pin to make a hole in its side near the bottom. Different diameters of each draining hole and two fluids that are water and cooking oil, were used to result the speed of efflux. A piece of sticky tape was put over the orifice and then filled the mixed-blue dye water to enhance visibility of water in the bottle. The orifice was opened to record the flow rate on video files by using a smartphone as shown in Fig.\ref{Setup}. Distance between the plastic bottle the camera is approximately 1 meter that is enough to ignore the distorted image of the image of a smartphone lens \cite{distortion}. Decreasing water level in the container analyzed using open-source software ‘Tracker’. For instance, decreasing oil level as a function of time in the container was determined by Tracker as shown in Fig.\ref{Tracker analysis}. We used Eq.({\ref{str5}}) to determine $a=0.5 g \lambda^2$, $b=-\sqrt{2g\lambda^2 H_{0}}$, and $H_{0}$. All results are shown in Tab.\ref{Tab1} and \ref{Tab2}.

\section{Result \& Discussion} \label{s4}
\begin{figure}[ht]
\begin{center}\includegraphics[width=1\linewidth]{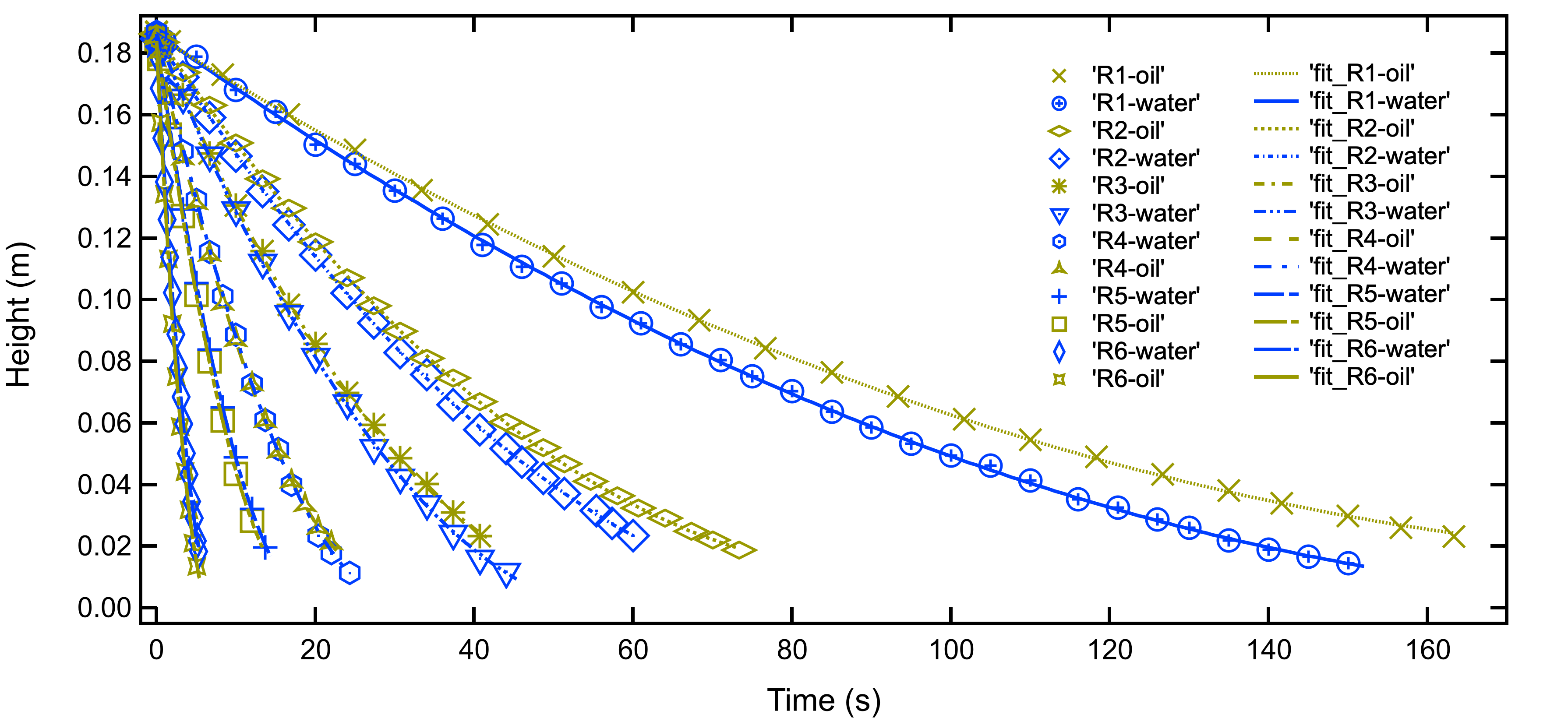}
\caption{We show a flow rate of water and oil obtained by varying a height of a hole-size, $Ri\,,i=1,\,2,\,3,\,4,\,5,\text{ and } 6$. We introduce six different hole sizes of the container.} 
\label{Results1}
\end{center}
\end{figure}
\begin{figure}[ht]
\begin{center}\includegraphics[width=0.85\linewidth]{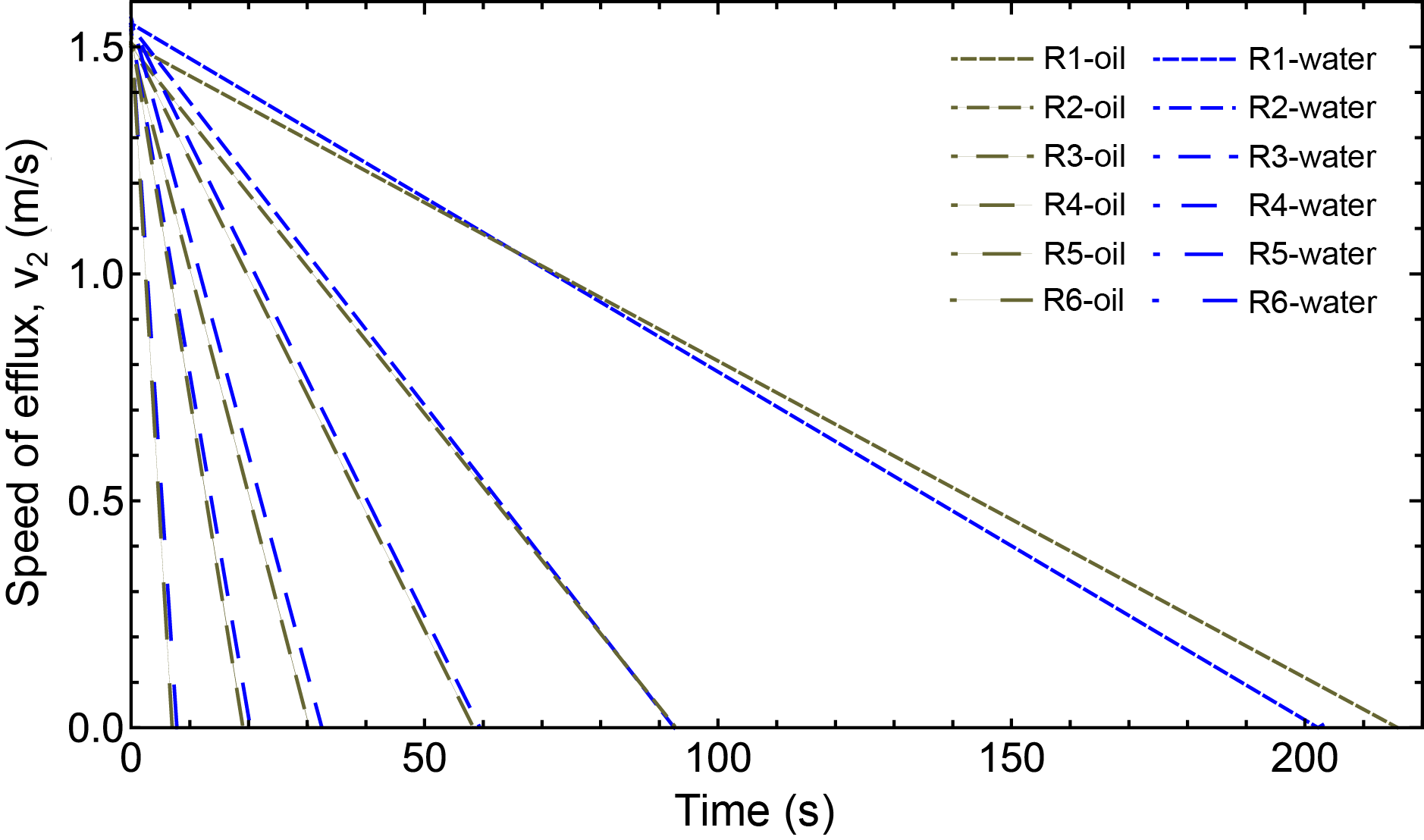}
\caption{We plot the speed of efflux of oil and water versus time when the size of  draining holes are varied.} 
\label{Speed ef}
\end{center}
\end{figure}

After tracking times of drain ($t$) and height ($H$) of the liquids by using Tracker, height-time graph of different hole-sizes was plotted as shown in Fig~\ref{Results1}. In every test, each liquid was filled up to the level of big cross-sectional area $A_1$ at about 18.5 cm, above the small cross-sectional area (orifice) $A_2$, as displayed Fig~\ref{botle Fig}. The area $A_1$ opens to atmospheric pressure. Once liquids were released from $A_2$ to the air, their levels non-linearly decreased before stopping at level of $A_2$. In our experiment, we have observed different draining time of the liquids from individual hole-sizes. In Fig~\ref{Results1}, draining time of water and oil at particular hole-sizes from inertial level are approximately the same, before splitting away from each other at about $H = 15$\,cm. The splitting of draining time can be observed from the orifices R1, R2 and R3 of radii 0.16, 0.22 and 0.28 cm, respectively. In case of bigger radii 0.40, 0.52 and 0.83 cm of the orifices R4, R5 and R6, water is comparable to oil along the course of draining.

Gradient of the height-time graph (Fig~\ref{Results1}) was computed and translated the graph into acceleration of leak, as shown in Fig~\ref{Speed ef}. Due to a density of water higher than that of oil, the high pressure formulates a higher speed of efflux that can be observed at the early time of every hole-size. Each liquid flows from the highest level to the lower one resulting in a dropping of pressure, and then the reducing of a speed. This results are in good agreement with the work done by Sianoudis and Drakaki \cite{undergrad}. We consider a case of viscosity and resistance of the fluid taking into account. At the smallest hole $R1$, after $60\,s$, as a more oil left in the bottle, pressure of the oil is higher and therefore it flows faster than the water. For instant at $63\,\text{second}$, pressure of the oil is $2$ Pascal higher than that of the water. Nevertheless, the oil to drain off after the water. According to effects of high viscosity of the oil--friction among its molecules and friction between its molecules and the hole \cite{slowdrain}, it converts some amount of kinetic energy into thermal energy. Consequently, the water flows with a higher deceleration when compare to the oil. These effects are dominantly present to the three smallest holes $R1,\,R2$ and $R3$. In case of $R4,\,R5$ and $R6$, their radii are sufficiently large to disregard friction due to hole-size effect \cite{Frank}. 

 The fit of parabolic function to height-time graph (Fig~\ref{Results1}) allows us to parametrize $a$ and $b$ which are macroscopic quantities that contain constant terms, $\lambda$, and $H_0$ in Eq.(\ref{str6}). Parameters $a$ and $b$ in Table \ref{Tab1} and \ref{Tab2} indicate that acceleration of the free surface ($A_1$) increases with the hole-size, while initial velocity of the free space decreases. In this work, radius of the holes $R_{\rm ex}$, initial height of liquids $H_0$ at $t=0$ and terminal time of draining are also verified. Since $a=0.5g\lambda^2=0.5g (A_{2}/A_{1})^2=0.5g ({\cal R}_{2}/{{\cal R}_{1}})^4$, ${\cal R}_{2}={\cal R}_{1}(2a/g)^{1/4}$, substitution of radius ${\cal R}_{1}=4.44\text{ cm}$ gives radius of the hole. Radius $R_{\rm ex}$ obtained from calculation is smaller than the radius $R_{\rm meas}$ measured by using vernier caliper with less than $20$ percent of difference. The difference is possibly due to uneven circumference of the hole occurred from burning in piecing process. The initial height of liquids $H_0$ obtains from $b^{2}{{\cal R}_{1}^{4}/2g{\cal R}_{2}^{4}}$, or from the last term in Eq.(\ref{str6}). The initial height of liquid from our experiment differs from our measurement less than 1 percent of difference. Regarding to fit data, the zero gradient provides us drain off time. The bigger the hole is, the faster liquid will reach its terminal time.    
 
\begin{table}[h]
 \caption{We show water draining through a hole using different hole sizes. We obtain best-fit parameters shown in the table. A terminal time can be estimated.}
\begin{tabular}{|c|c|c|c|c|c|c|}
\hline
Hole Type & $R_{\rm meas}$ ($cm$) & $a\,(m/s^{2})$        & $b\,(m/s)$         & $R_{\rm ex}$ ($cm$) & $H_{0}\,(cm)$ & Terminal time ($s$) \\ \hline
R1 & 0.1634$\pm$0.012         & 4.50E-06 & -1.82E-03 & 0.14$\pm$0.03 & 18.64       & 202.44            \\ \hline
R2 & 0.229$\pm$0.016        & 2.15E-05 & -3.98E-03 & 0.20$\pm$0.02 & 18.50        & 92.56             \\ \hline
R3 & 0.277$\pm$0.011         & 5.30E-05 & -6.30E-03 & 0.25$\pm$0.06 & 18.61       & 59.43             \\ \hline
R4 & 0.402$\pm$0.017       & 1.80E-04 & -1.17E-02 & 0.35$\pm$0.06 & 18.62      & 32.42             \\ \hline
R5 & 0.520$\pm$0.012         & 4.43E-04 & -1.80E-02 & 0.43$\pm$0.02 & 18.55     & 20.32             \\ \hline
R6 & 0.834$\pm$0.014         & 3.01E-03 & -4.69E-02 & 0.70$\pm$0.04 & 18.35       & 7.79              \\ \hline
\end{tabular}
\label{Tab1}
\end{table}

\begin{table}[h]
 \caption{We show oil draining through a hole using different hole sizes. We obtain best-fit parameters shown in the table. A terminal time, a time of water that starts draining to that stops draining, can be estimated.}
\begin{tabular}{|c|c|c|c|c|c|c|}
\hline
Hole Type & $R_{\rm meas}$ ($cm$) & $a\,(m/s^{2})$        & $b\,(m/s)$         & $R_{\rm ex}$ ($cm$) & $H_{0}\,(cm)$ & Terminal time ($s$) \\ \hline
R1 & 0.163$\pm$0.012         & 3.80E-06               & -1.64E-03   & 0.13$\pm$0.03                   & 18.58                      & 215.79            \\ \hline
R2 & 0.229$\pm$0.016         & 2.02E-05               & -3.75E-03   & 0.20$\pm$0.01                 & 18.60                       & 92.82            \\ \hline
R3 & 0.277$\pm$0.016         & 5.22E-05               & -6.09E-03             & 0.25$\mp$0.06     & 18.60                        & 58.33             \\ \hline
R4 & 0.402$\pm$0.012          & 1.91E-04               & -1.16E-02             & 0.35$\pm$0.02     & 18.33                       & 30.39             \\ \hline
R5 & 0.520$\pm$0.012         & 4.96E-04               & -1.89E-02             & 0.45$\pm$0.06    & 18.41                        & 19.09             \\ \hline
R6 & 0.834$\pm$0.014         & 3.75E-03               & -5.24E-02             & 0.74$\pm$0.05     & 18.34                        & 6.99              \\ \hline
\end{tabular}
\label{Tab2}
\end{table}

\section{Conclusion}
In this work, we have performed experimental investigation of liquid draining through a hole of a container. We developed this experiment with intention to propose it to be a lab at home. The use of equipment commonly available at home or that can be purchased at a low price. We have introduced a freely-accessible video analysis software , Tracker, to assist us in a process of data analysis to obtain accurate results. 

Interestingly, we observed effects of viscosity causing a delay of draining time of liquids. Resistant force according to the viscosity converts some kinetic energy of moving liquid into thermal energy resulting in a reduction of flow speed \cite{Frank}. Our study showed that a ratio of the orifice radius to the free surface radius $({\cal R}_{2}/{\cal R}_{1})$ can be used to determine a terminal time. The greater ratio is, the slower terminal time is obtained. Moreover, we have found a proportional relation of $({\cal R}_{2}/{\cal R}_{1})^4$ to acceleration of declining level of liquid in the container. These behaviors are corresponding to Poiseuille's law, and in a good agreement with a previous study \cite{undergrad}. However, the effect of hole size has been only observed for small holes.  

Following fluid dynamics principles and a mathematical function, we obtained Eq.(\ref{str6}) written in a standard parabolic form. This measures a change of liquid level as a function of time. Disclosure of coefficients of parabolic function fits to our experimental data demonstrating that a quadratic coefficient "$a$" and a linear one "$b$" are interpreted as an acceleration and an initial velocity of liquid in the container, respectively. The terms of a ratio $({\cal R}_{2}/{\cal R}_{1})^4$ and an initial level of liquid in the container are concealed in the coefficients determining the instantaneous height of liquid left in the container. In addition, following our experimental approach, ones can examine a hole-size, quantify an initial height of liquid in a container, and estimate the terminal time of draining in actual situation.

Our experiment related to fluid mechanics possibly implements in fundamental physics laboratory course for freshmen either in engineering or in science program. We believe that this experimental template integrated with computer assisted data analysis software will valuable to various educational aspects, including; 1) improvement for increase students' hands-on involvement. Experience with hands-on experiment through trial and error allows the students observe physics phenomena, learn from mistakes, and distinguish gap between theory and practice, 2) enhancement of students' understanding in basic fluid mechanics principles to be able to apply the knowledge for solving practical problems, 3) adaptation of using Tracker software enabling students or any experimenters to accurately analyze moving objects in other experiments, such as tracking flow of fluid to observe velocity distribution across and along a pipe, tracking trajectory of projectile, 4) elimination of fund limitation to purchase practical experimental equipment and barriers to distance learning, and 5) advancement of STEM/STEAM education to students. Our proposed laboratory enables students to transfer their knowledge across disciplines that are physics, engineering, mathematics, technological tool of analysis software, and art in the experiment. Last but not least, our proposed scenario, based on a lab@home strategy, constitutes the doable template for teachers/students to setup, perform, and analyse other physics experiments at home.


\appendix

\section{Lab@home guide}
The proposed lab of fluid draining contains 3 processes; preparation, recording, and analyzing. All the processes consume about 3 hours. In preparation process, according to our pilot study, we would suggest an experimenter to use at least 3 transparent plastic bottles of the same shape and dimension. They can be reused ones. Discrete hole sizes can be made by pressing different sizes of hot metal pins onto the plastic bottles, and moving the tool around into a circle to get rid of uneven edge. Fluids such as water, cooking oil, salt water, glycerine, etc.can be applied in this experiment. To enhance visibility of water level, a small drop of dye can be dissolved into it. Video record can be achieved by using a camera from either a smartphone, or a professional one. A stable smartphone or camera holder such as a tripod, table, chair, etc. and a bright room are recommended for obtaining a good quality VDO shot. The last thing to prepare is a computer (or laptop/tablet) with installation of Tracker software from https://physlets.org/tracker/ and spreadsheet application. Besides, the Tracker software is possible to run online. All in this part would take about an hour to get ready for the next part of tracking.    

Tracker software requires an input of frame rate and a visible known length in a recorded VDO frame. The frame rate can be found in device specific data. In case of unknown frame rate, the experimenter may test it via online tool available on the Internet. In laboratory, frame rate can be counted by using stroboscope. The known length as a ruler is suggested to present in VDO recording for calibration of distance. During flowing out of the small hole, height of fluid levels have to be tracked at certain times. Tracking point should not less than 20 points distributed through out the flow, equivalent to 5 step size. Process of tracking data might consume an hour. 

Data obtained from Tracker software feasibly export or copy to a spreadsheet application such as Excel, Number, Google Sheet, etc., enabling the experimenter to calculate, plot and fit the data. Height-Time graphs of particular flowing liquids out of varied hole sizes assist the experimenter to find drain off time and parametrize $a$ and $b$ for determination of initial height of water in the bottle, $H_0$ and size of the leakage $R_{ex}$. Gradient of the height-time graphs is interpreted as speed of efflux-time graph analogous to liquid acceleration of the flow. The best-fit parabola to height-time data of the form $at^2+bt+c$, given in Eq.(\ref{str6}), yields $a$ and $b$ coefficients to quantify $H_0$ and ratio of open surfaces $A_2$ to $A_1$ from $\lambda$, respectively. With a known area, the experimenter is able to compute another one. In process of data analysis, the experimenter would spend about an hour in our proposed lab.


\begin{thebibliography}{99}

\bibitem{labathome}
E. G. Campari, M. Barbetta, S. Braibant, N. Cuzzuol, A. Gesuato, L. Maggiore, F. Marulli, G. Venturoli, and C. Vignali, 2021. Physics Laboratory at Home during the Covid-19 Pandemic, The Physics Teacher, 59(1), 68–71.
\bibitem{Serway}
A. Serway, R. and W. Jewett, Jr., J., 2008. Physics for Scientists and Engineers with Modern Physics. 9th ed. Boston: Mary Finch, Charlie Hartford.
\bibitem{distortion}
B. Colton, W. Alison, D.Anna, C. Jenna, W. Margaret, 2021. Seeing Is Not Believing: Facial Distortion in Smartphone Photography, Plastic \& Reconstructive Surgery.
\bibitem{undergrad}
A. Sianoudis, E. Drakaki, 2008. An Approach to Poiseuille's Law in an Undergraduate Laboratory Experiment, European Journal of Physics, 29, 489 - 495.
\bibitem{slowdrain}
J. N. Libii, 2002. Mechanics of the Slow Draining of a Large Tank under Gravity. American Journal of Physics, 71(11), 1204-1207.
\bibitem{Frank}
F. M. White. Fluid Mechanics, 8th edition, McGraw-Hill Education, 2016.





\end{thebibliography}
\end{document}